\documentclass{evn2024}
\usepackage{graphicx}
\usepackage{url}
\usepackage{amsmath}
\usepackage{amssymb}
\usepackage{natbib}
\usepackage{hyperref}
\begin{document}
   \title{Dynamical Masses of Young Stellar Multiple Systems with the VLBA (DYNAMO-VLBA)}

   \author{Jazmín Ordóñez-Toro\inst{1}\thanks{E-mail: n.ordonez@irya.unam.mx},
          Sergio A. Dzib\inst{2}
          \and
          Laurent Loinard\inst{1,2,3}
          }

   \institute{Instituto de Radioastronom\'{\i}a y Astrof\'{\i}sica, Universidad Nacional Aut\'onoma de M\'exico, Apartado Postal 72-3, Morelia 58089, M\'exico
         \and
             Max Planck Institut f\"ur Radioastronomie, Auf dem Hügel 69, D-53121 Bonn, Germany
          \and 
             Black Hole Initiative at Harvard University, 20 Garden Street, Cambridge, MA 02138, USA
           \and
           David Rockefeller Center for Latin American Studies, Harvard University, 1730 Cambridge Street, Cambridge, MA 02138, USA
             }

\authorrunning{Ordóñez-Toro et al.}
\titlerunning{DYNAMO-VLBA dynamical masses of YSOs}

   \abstract{
Very Long Baseline Interferometry (VLBI) provides high angular resolution images and has been used for stellar astrometry for decades. The DYNAMO-VLBA project utilizes the Very Long Baseline Array (VLBA) to study tight binary and multiple pre-main sequence stars, whose components have detectable radio emission and typical separations on the order of milli-arcseconds. Such systems cannot be resolved by Gaia, making VLBI an essential tool for the study of their orbital parameters and, eventually, the determination of their mass. Here, we report VLBA dynamical mass measurements of the individual stars in the S1 system in Ophiuchus and EC\,95 in Serpens. S1 is the most luminous and massive stellar member of the nearby Ophiuchus star-forming region. We find that the primary component, S1A, has a mass of $4.11 \pm 0.10\,M_{\odot}$. This is significantly less than the value of $\sim6\,M_{\odot}$ expected from theoretical models given the location of S1A on the HR diagram. The secondary, S1B, has a mass of $0.831 \pm 0.014\,M_{\odot}$ and is most likely a T Tauri star. In the Serpens triple system EC\,95, we measure the masses of EC\,95A and EC\,95B, finding $2.15\pm0.10$ M$_\odot$ and $2.00\pm0.12$ M$_\odot$, respectively. In this case, the measured masses agree with the location of the stars in the HR diagram for very young 2 $M_\odot$ stars. For the first time, we also estimated the mass of tertiary, EC\,95C, to be 0.26 $^{+0.53}_{-0.46}$ M$_\odot$.% The dynamical mass estimations derived from VLBA data are free of assumptions on the physical parameters of the stars and could be used to test evolutionary models of pre-main sequence stars.
   }

   \maketitle
%
%________________________________________________________________

\section{Introduction}
The precise measurement of stellar masses is fundamental for understanding star formation processes and stellar evolution. Young pre-main sequence stars in nearby regions offer a unique opportunity to study these processes in detail. However, determining the masses of these stars is challenging, especially when dealing with binary or multiple systems with very small separations. Very Long Baseline Interferometry (VLBI) provides high angular resolution images and has been used for stellar astrometry for decades. In particular, the \textit{Very Long Baseline Array} (VLBA) is an essential tool for studying stellar systems with separations on the order of milliarcseconds, where space missions like \textit{Gaia} cannot resolve individual components or provide information on their orbital parameters.\\

In the \textit{Dynamical Masses of Young Stellar Multiple Systems with the VLBA} (DYNAMO-VLBA)\footnote{\url{https://www3.mpifr-bonn.mpg.de/div/radiobinaries/intro.php}} project, we use the VLBA to study binary and multiple pre-main sequence stars whose components have detectable radio emission. These observations allow us to measure the dynamical masses of individual stars without relying on assumptions about their physical parameters. This is crucial for testing and refining theoretical evolutionary models of young stars.\\

One of the systems of particular interest is S1 in the Ophiuchus star-forming region. Located at a distance of $137.2 \pm 0.4$\,pc \citep{Ordonez-Toro2024}, S1 stands out as the brightest and most massive stellar member of this region. It was the first young stellar object directly detected using VLBI techniques \citep{andre1991}. Lunar occultation experiments in the infrared \citep{richichi94} and radio observations with the VLBA \citep{ortiz2017a} revealed that S1 is a binary stellar system with an angular separation on the order of 20\,mas. Previous studies suggested that S1 is a young Herbig Be star with an estimated mass of 5 to $6\,M_{\odot}$ based on photometric measurements \citep[e.g.,][]{lada1984}.\\

Another system of interest is EC\,95 in the Serpens region, located at $436.0 \pm 9.2$\,pc \citep{ortiz2018b}. EC\,95 was initially classified as a proto-Herbig Ae/Be star of spectral type K2, with an approximate age of $10^5$ years and an estimated mass of $\sim4\,M_{\odot}$ based on its location in the HR diagram \citep{Preibisch1999}. Radio observations with the VLBA revealed that EC\,95 is a binary system with a separation of approximately 15\,mas \citep{dzib2010}. These observations indicated that the system consists of two components, EC\,95A and EC\,95B, and suggested a significantly higher mass for the primary. Additionally, a third component, EC\,95C, was detected in near-infrared observations with the VLT \citep{Duchene2007} and in radio observations with the VLBA \citep{ortiz2017b}, making EC\,95 a hierarchical triple system where each component shows non-thermal radio emission.

%__________________________________________________________________
\section{Observations and Data Reduction}
Observations of the stellar systems S1 and EC\,95 were carried out using the VLBA of the National Radio Astronomy Observatory (NRAO) as part of the DYNAMO-VLBA project (VLBA project code: BD215). The observations recorded the radio continuum flux at a wavelength of $\lambda = 6.0$\,cm ($\nu = 5$\,GHz).
The data calibration followed standard procedures for phase referenced VLBI observations and was performed using the Astronomical Image Processing System (\textsc{AIPS}) software \citep{Greisen2003}. These procedures have been detailed in previous works \citep[e.g.,][]{loinard2007,dzib2010,ortiz2017a}. Flux densities and positions of detected sources were measured using a two-dimensional Gaussian fitting procedure (task \texttt{JMFIT} in \textsc{AIPS}).

In addition to the DYNAMO-VLBA observations, we included archival VLBA observations of S1 and EC\,95 in our analysis. For S1, this includes data from the Gould Belt Distances Survey (GOBELINS) previously reported by \citet{loinard2008,ortiz2017a,ortiz2018b}. For EC\,95, archival data from projects by \citet{dzib2010}, and the GOBELINS survey were used \citep{ortiz2017b}. In total, we analyzed 35 epochs for S1 and 32 epochs for EC\,95, spanning more than a decade of observations for each source.

\section{Astrometric Fitting Procedure}
The motion of the components of a binary stellar system on the celestial sphere can be described by the combination of their common trigonometric parallax $\pi$, the uniform proper motion of their center of mass in right ascension ($\mu_\alpha$) and declination ($\mu_\delta$), and their orbital motions around the center of mass. For the primary component, the equations of motion are expressed as:
\begin{eqnarray}
\alpha(t) & = & \alpha_0 + \mu_\alpha t + \pi f_\alpha(t) + a_1 Q_\alpha(t),\label{eqn:pm1}\\
\delta(t) & = & \delta_0 + \mu_\delta t + \pi f_\delta(t) + a_1 Q_\delta(t),\label{eqn:pm2}
\end{eqnarray}
In these equations, $f_\alpha(t)$ and $f_\delta(t)$ correspond to the projections of the parallax ellipse in right ascension and declination, while $Q_\alpha(t)$ and $Q_\delta(t)$ represent the projections of the orbital motions. The detailed expressions are provided in \citet{Ordonez-Toro2024}. With appropriate minor adjustments, similar equations hold for the secondary component.

To estimate the astrometric and orbital parameters of the binary system, we follow the procedure described in \citet{Ordonez-Toro2024}. We applied the MPFIT least squares fitting algorithm \citep[see also][and references therein]{kounkel2017}, which fits the observed positions to the equations of motion. This approach allows us to determine the total mass of the system from Kepler's third law and the individual masses of the stars by accounting for their motions relative to the center of mass.

\section{Results}

\subsection{Binary system S1}
Our VLBA observations, combined with archival data, resulted in a total of 35 epochs. The primary component, S1A, was detected in all 35 epochs, while the secondary component, S1B, was detected in 14 epochs.

\begin{figure}
\includegraphics[width=0.45\textwidth]{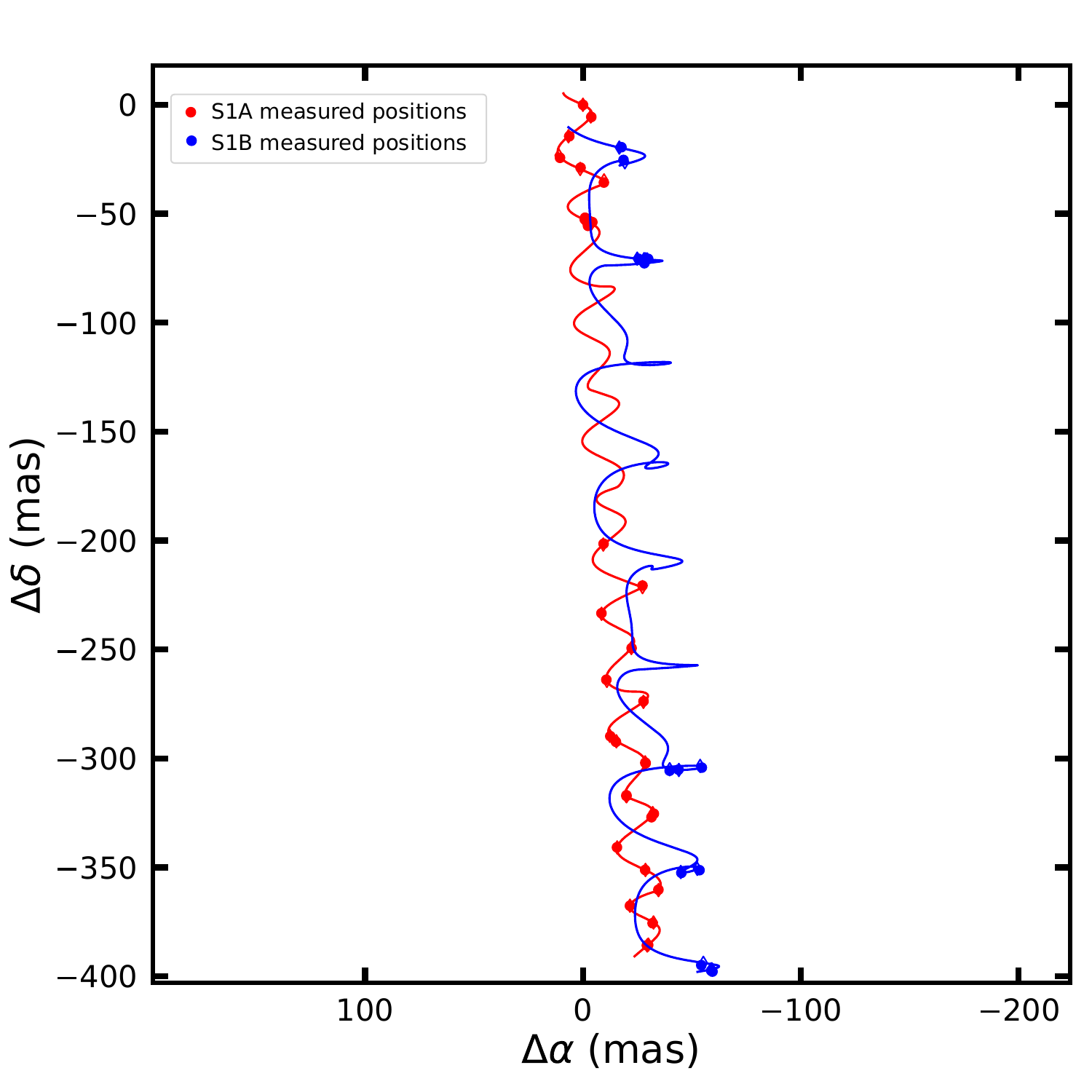}
\caption{Measured positions of S1A (red dots) and S1B (blue dots) shown as offsets from the position of S1A in the first detected epoch (2005 June 24).}
\label{s1_1}
\end{figure}

\begin{figure}
\includegraphics[width=0.45\textwidth]{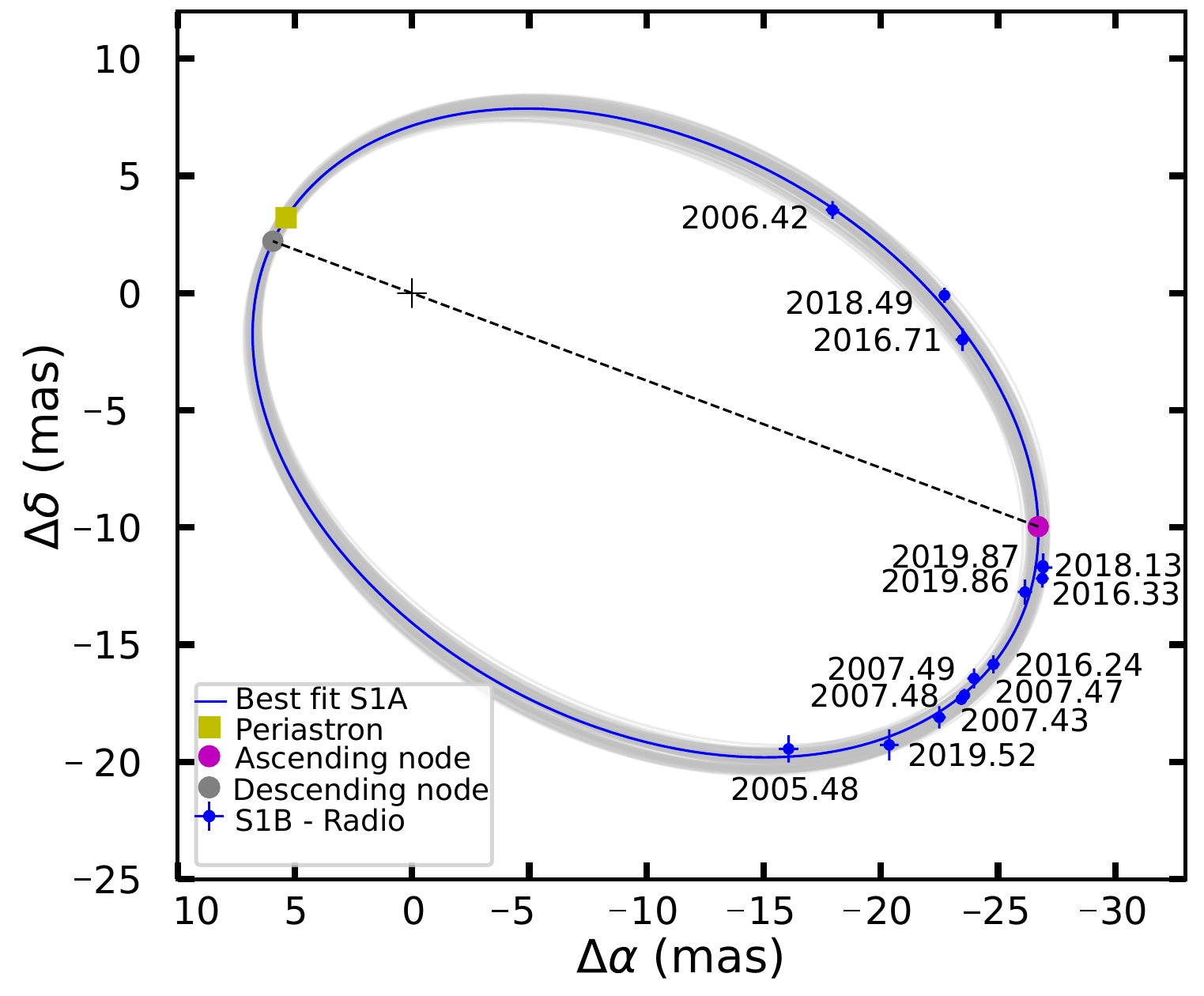}
\caption{Stellar relative positions and orbital fit model of S1. The blue dots indicate the relative positions of S1B with respect to S1A, and the errorbars consider the position errors of both components which are added in quadrature. The dashed black line traces the line of nodes from the model, and the black cross indicates the position of the primary.}
\label{s1_2}
\end{figure}

We determined that S1A has a mass of $4.11 \pm 0.10\,M_{\odot}$, significantly less than the previously reported $6\,M_{\odot}$. This highlights a discrepancy where models corresponding to the location of S1A on the HR diagram predict masses at least 25\% higher than the dynamical mass. The secondary, S1B, has a mass of $0.831 \pm 0.014\,M_{\odot}$, consistent with a low-mass star. Figure~\ref{s1_1} shows the measured positions of S1A and S1B. In Figure~\ref{s1_2}, we present the relative positions of S1B with respect to S1A, along with the orbital fit model.


\subsection{Binary system EC\,95}
For the EC\,95 system, our combined observations provided 32 epochs for EC\,95A and EC\,95B. The primary component, EC\,95A, was detected in all 32 epochs, and the secondary component, EC\,95B, was detected in 23 epochs.

We measured the masses of EC\,95A and EC\,95B to be $2.15 \pm 0.10\,M_{\odot}$ and $2.00 \pm 0.12\,M_{\odot}$, respectively. Additionally, we detected the third component, EC\,95C, combining four epochs of observations from infrared data \citep{Duchene2007} and VLBA measurements. For the first time, we estimated the mass of EC\,95C to be $0.26^{+0.53}_{-0.46}\,M_{\odot}$ with an orbital period of $172 \pm 14$ years.

\begin{figure}
\includegraphics[width=0.45\textwidth]{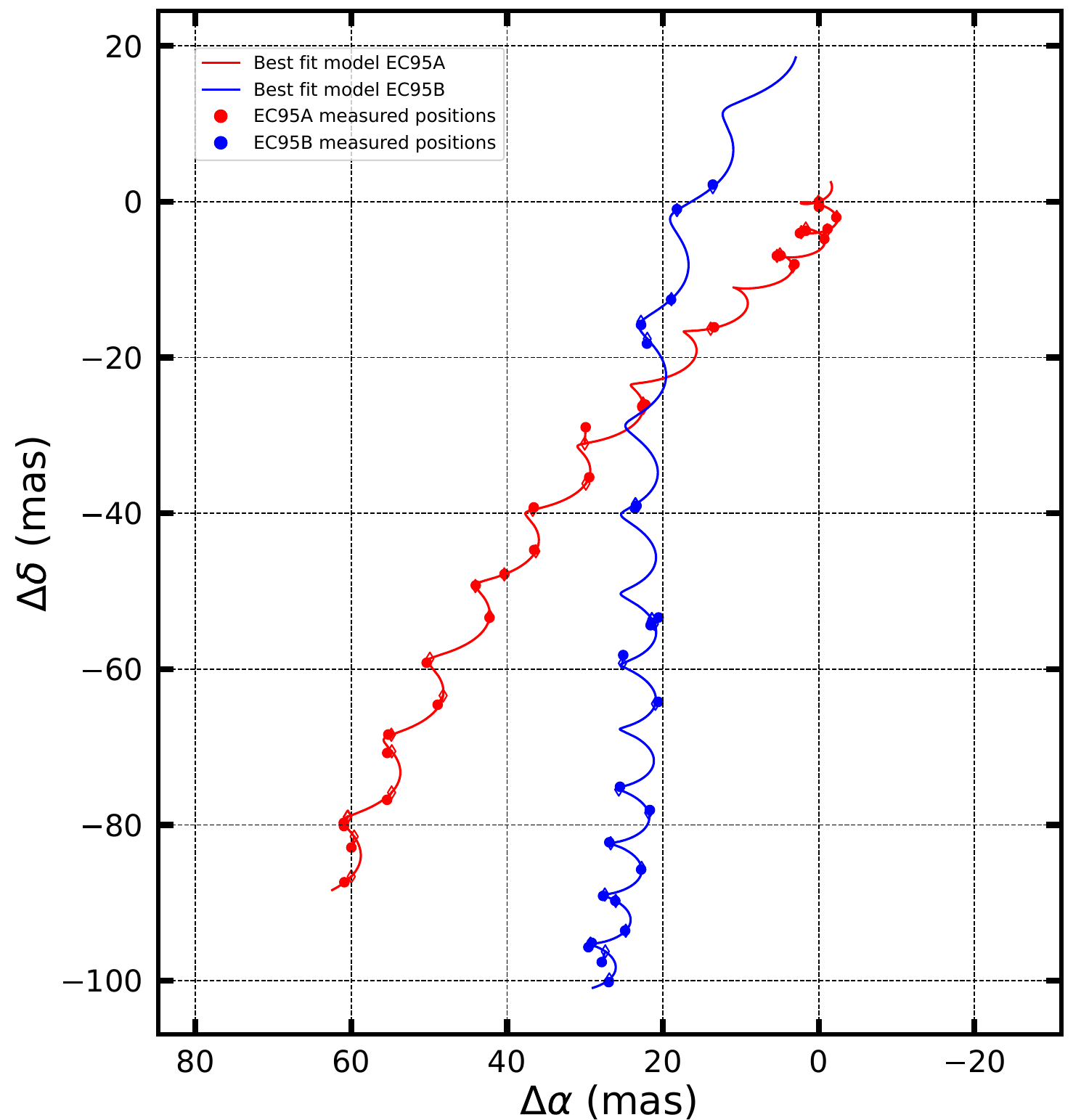}
\caption{Measured positions of EC\,95A (red dots) and EC\,95B (blue dots) shown as offsets from the position of EC\,95A in the first detected epoch (2007 December 22).}
\label{ec95_1}
\end{figure}

\begin{figure}
\includegraphics[width=0.45\textwidth]{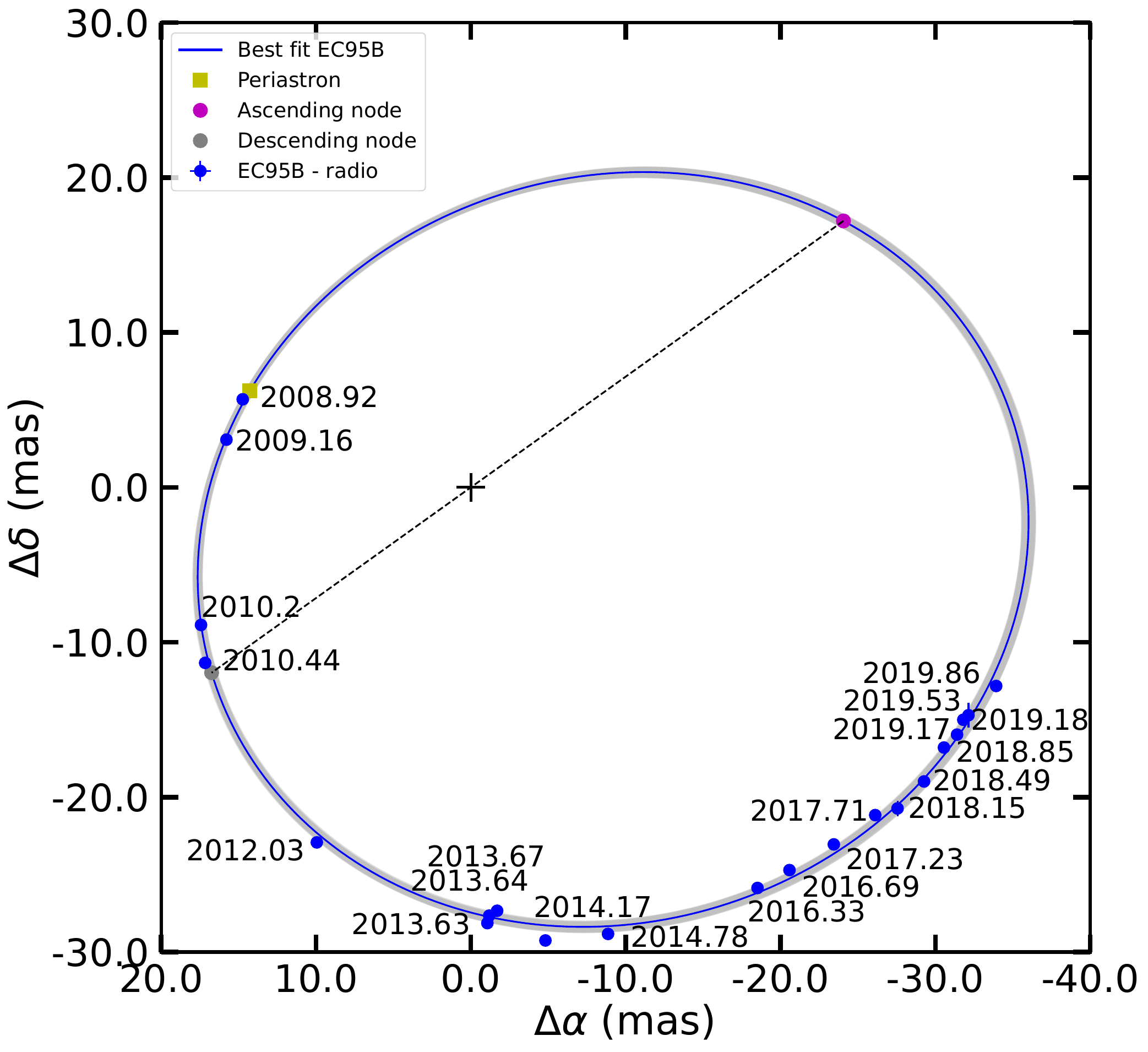}
\caption{Stellar relative positions and orbital fit model of EC\,95. The blue dots indicate the relative positions of EC\,95B with respect to EC\,95A. The remaining description is similar to that provided for the S1 Figure~\ref{s1_2}. }
\label{ec95_2}
\end{figure}

\begin{figure}
    \centering
    \includegraphics[width=0.45\textwidth]{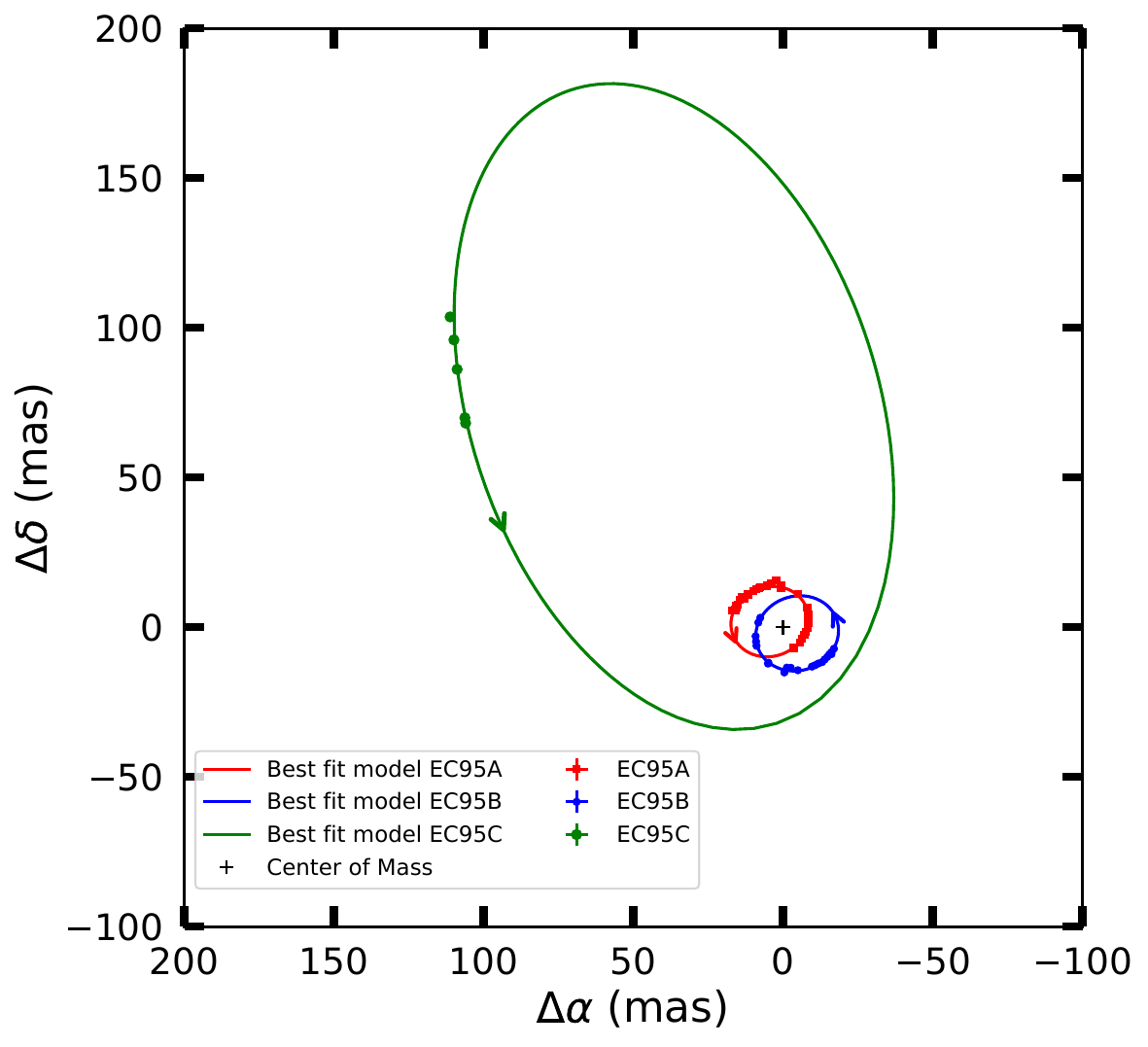}\hspace{0.2cm}
    \caption{Orbits of EC\,95A (red line), EC\,95B (blue line) and EC\,95C (green line) around the center of mass of the system (black cross). The colored squares indicate the measured positions, while the arrows show the direction of the orbits.}
    \label{Ec95abc}
\end{figure}

Figure~\ref{ec95_1} displays the measured positions of EC\,95A and EC\,95B. In Figure~\ref{ec95_2}, we present the relative positions of EC\,95B with respect to EC\,95A, along with our orbital fit model. Furthermore, Figure~\ref{Ec95abc} shows the orbital configuration of the entire system, including the components A, B, and C around the center of mass. 


%______________________________________________________________

\section{Discussion and Conclusions}
In this work, we presented new VLBA observations of the S1 and EC\,95 systems, combined with archival data, to improve the constraints on the dynamical masses of the individual stars in these young stellar multiple systems.\\

For the S1 system, our analysis determined that the primary component, S1A, has a dynamical mass significantly less than previous estimates based on photometric measurements \citep[e.g.,][]{lada1984}. This discrepancy shows that theoretical models corresponding to the location of S1A on the HR diagram overpredict the mass by at least 25\%. By considering models that incorporate rotation (specifically the PARSEC pre-main sequence models of \citealt{parsec2022}), we show that rotation does not reconcile the HR diagram position of S1A with its dynamical mass measurement \citep{Ordonez-Toro2024}. Therefore, refinements in the evolutionary models for young intermediate-mass stars like S1A are necessary to reconcile these differences. The secondary component, S1B, is consistent with a low-mass T Tauri star.\\

For the EC\,95 system, we found that the dynamical masses of the primary and secondary components are consistent with each other and with early stellar evolution models. This agreement supports the reliability of current pre-main sequence evolutionary tracks for low-mass stars. Our detection of the third component in the EC\,95 system (EC\,95C) and the constraints we obtain on its orbit also enable us to discuss the arquitecture of the entire system. We find that the inclinations of the inner binary orbit (AB) and the orbit of EC\,95C around the barycenter of AB are both approximately $35^\circ$, which is consistent with a hierarchical system formed through disk fragmentation \citep{Adams1989}. However, the significant eccentricities of both orbits ($e = 0.391 \pm 0.003$ for AB and $e = 0.75 \pm 0.03$ for EC\,95C) and the misalignment of their semi-major axes raise questions about the hierarchical stability of the system. The orientations of the semi-major axes, with $\omega = 117.27 \pm 0.51$ degrees and $\Omega = 305.56 \pm 1.13$ degrees for AB, and $\omega = 43.49^{+11.03}_{-7.34}$ degrees and $\Omega = 169.23^{+7.71}_{-12.44}$ degrees for EC\,95C, further emphasize these challenges. As seen in Figure~\ref{Ec95abc}, near periastron, the separation between EC\,95C and the AB barycenter becomes comparable to the separation between EC\,95A and EC\,95B, potentially leading to dynamical instability. It is important to note that the limited orbital coverage of EC\,95C introduces uncertainties in its orbital elements. Future resolved observations at radio and infrared wavelengths will be crucial to assess the long-term stability of the system.\\

Overall, our study emphasizes the importance of precise dynamical mass measurements in young close binary systems and demonstrates the effectiveness of VLBA observations in achieving this goal. These results help to constrain pre-main sequence stellar evolution models and improve our understanding of stellar formation and evolution in multiple systems.

\begin{acknowledgements}

J.O. and L.L. acknowledge the financial support of CONAHCyT, M\'exico.
L.L. acknowledges the support of DGAPA PAPIIT grants IN112416, IN108324 and IN112820 as well as CONACyT-CF grant 263356.
%L.L. acknowledges the financial support of DGAPA, UNAM (project IN112417).
S.A.D. acknowledge the M2FINDERS project from the European Research Council (ERC) under the European Union's Horizon 2020 research and innovation programme (grant No 101018682).
The National Radio Astronomy Observatory is a facility of the National Science Foundation operated under cooperative agreement by Associated Universities, Inc.

\end{acknowledgements}

\end{document}